\documentclass[aps,prl,superscriptaddress,twocolumn,showpacs,showkeys]{revtex4}

\usepackage{amsmath}
\usepackage{amssymb}
\usepackage{amsfonts}
\usepackage{graphicx}

\newcommand{\dd}{\textrm{d}}
\newcommand{\cotan}{\textrm{cotan}\,}
\newcommand{\Lie}{{\cal L}}

\newcommand{\UD}[2]{\ensuremath{^{#1}_{\phantom{#1} #2}}}

\newenvironment{definition}[1][Definition]{\begin{trivlist}
\item[\hskip \labelsep {\bfseries #1}]}{\end{trivlist}}

\newcommand{\qed}{\nobreak \ifvmode \relax \else
      \ifdim\lastskip<1.5em \hskip-\lastskip
      \hskip1.5em plus0em minus0.5em \fi \nobreak
      \vrule height0.75em width0.5em depth0.25em\fi}

\begin{document}

\title{Quasi--local angular momentum of non--symmetric isolated and dynamical horizons
from the conformal decomposition of the metric}

\author{Miko\l{}aj Korzy\'nski}
\email[]{mkorz@fuw.edu.pl}

\affiliation{Institute of Theoretical Physics, Warsaw University, ul. Hoza 69, 00-681 Warszawa, Poland}
\affiliation{Max--Planck--Institut f\"ur Gravitationsphysik (Albert--Einstein--Institut), Am M\"uhlenberg 1, 14476 Golm, Germany}


\begin{abstract}
A new definition of quasi--local angular momentum of non--axisymmetric
marginally outer trapped surfaces is proposed. It is based on conformal 
decomposition of the two--dimensional metric and the action
of the group of conformal symmetries. The definition is completely general and agrees with the standard one in axisymmetric surfaces. 
\end{abstract}

\maketitle

\section{Introduction}
The  quasi--local theory of black hole boundaries is based on the notion of
marginally outer trapped surfaces (MOTS's), i.e. two--dimensional spacelike
surfaces of spherical topology, for which the expansion of one of the 
null normals, say $l^\mu$, vanishes. The tubes made of such surfaces are called
isolated, dynamical, trapping or slowly evolving horizons depending
on a number of additional assumptions \cite{ashtekar-2002-89, booth-2004-92, korzynski-2006-74, hayward-1994}. Although the theory is now at a mature stage and has already found a broad scope of applications \cite{ashtekar-2004-7}, there remains an important gap in the formalism. Namely, there is no
unique definition of black hole angular momentum on horizons which 
do not admit any symmetries. Definitions based on approximate symmetry vectors
either fail in certain special cases, like the one presented in \cite{hayward-2006-74}, or 
 involve solving variational problems \cite{cook-2007}.
Lack of a universal definition of angular momentum $J$ is an important caveat, because the Smarr's formula,
which is the best candidate for the quasi--local black hole mass definition \cite{ashtekar-2001-64}, explicitly involves the value of $J$.

Mariginally outer trapped surfaces are equipped with a positive definite metric tensor which we will denote by
$q$, an area form $\epsilon$ compatible with the metric, and the rotation one form, 
which is the pullback of the derivative of one null normal contracted with the other one
\begin{equation}
\omega_A = -(\nabla_A l^\mu)\,k_\mu = (\nabla_A k^\mu)\,l_\mu
\end{equation}
with the normalization condition $l^\mu\,k_\mu = -1$ imposed.
(The capital Latin index denotes here a geometric object defined on the MOTS as
opposed to spacetime objects for which Greek indices are used.)
We note here that contrary to the metric, the rotation form is not gauge invariant:
 any renormalization
of the null normals $l^\mu \to C\,l^\mu$, $k^\mu \to C^{-1}\,k^\mu$ and, in case of an isolated horizon, any change of the 
horizon foliation results
in adding a gradient to the rotation form \cite{korzynski-2006-74,lewandowski-2006-23}. 

All definitions of quasi--local angular momentum of 
a MOTS, denoted here by $\Delta$, are equivalent to a single integral formula
\begin{equation}
J_\phi = -\frac{1}{8\pi G} \int_\Delta \omega(\phi)\,\epsilon \label{eqj}
\end{equation}
involving a vector field $\phi$ on $\Delta$  (for a review see \cite{hayward-2006-74}).

In the Hamiltonian formalism every expression of this type  is the 
Hamiltonian generator of
diffeomorphisms generated by $\phi$ \cite{booth-2005-22}. In physics, however, we associate angular momentum only with
rotations and
if $\phi$ is supposed to generate anything of that type, it
must be an axial vector, i.e. have exactly two poles, closed integral curves and be normalized in
such a way that each integral curve closes when the affine parameter attains $2\pi$.

If the MOTS admits an axial symmetry, there exists a Killing vector
field $X$ of desired properties which one can plug into (\ref{eqj}). 
However if no such vector exists, we are left with a wide range of possible choices, and
consequently with a large room for arbitrariness. 
In fact, the freedom of choice is governed by the whole group of diffeomorphisms of $\Delta$. 
Thus the problem of assigning angular momentum to a non--symmetric isolated or
dynamical horizon is in fact the problem of picking up an appropriate vector field which one would substitute in (\ref{eqj}), rather than deriving from scratch an expression for $J$.

We postulate that a reasonable proposition for the choice for $\phi$ and $J$ should satisfy 
four mild, physically motivated conditions: 
\begin{enumerate}
\item $\phi$ and $J$ must be defined unambiguously for all possible 
horizon geometries, with or without symmetries,
\item $\phi$ must always be an axial vector in the sense defined above,
\item if the MOTS admits an axial symmetry vector $X$ of both metric $q$ and one--form $\omega$, 
the prescribed $\phi$ must agree with $X$ and $J$ with $J_X$, perhaps up to sign, 
\item the definition should be simple and natural, involving only the 
 two main ingredients of the MOTS geometry, i.e. $q$ and $\omega$.
\end{enumerate}
The first condition demands that we provide a construction that always works.
The third, among other things, assures that the definition will yield
the expected result when applied to a Schwarzschild or Kerr--Newman black hole
boundary.
The fourth one is to some extent an aesthetic requirement, but it is nevertheless 
 useful as a guiding principle in what is essentially an open problem, where many possible solutions exist.

In this paper we will provide a general formula for $J$ which satisfies all these conditions. We will  describe a construction of
the corresponding axial vector $\phi$, which works for all
horizons except a rather narrow class.
We will also provide a way to fix the gauge freedom
of the rotation form using the Hodge decomposition. This gauge
fixing removes the last ambiguity in the angular momentum definition and,
among other things, ensures that the expression for $J$ will yield the correct value
 in Kerr horizons even if
we choose a non--standard, tilted MOTS.  

The mathematical framework of the paper is based on a global decomposition
of the metric to the ``round'' spherical metric and the conformal factor.
This kind of decomposition has been used in the context of the black hole
boundaries \cite{czuchry-2004-70, caudill-2006-74}, though without any formal justification
of the method or discussing its invariance.
In \cite{czuchry-2004-70} it is ony applied to axisymmetric black holes and without any
attempts of providing an invariant definition of angular momentum. 
The authors of \cite{caudill-2006-74} on the other hand, while constructing 
 initial data for binary black holes using the conformal thin--sandwich decomposition, 
 make use of the three 
 Killing vectors
of the flat conformal metric on the time slice to define a measure
the spin of a black hole. Their method is equivalent to ours in many special cases, but not in general. The dependence on the choice of the conformal decomposition is not considered 
in their paper and therefore their prescription, when applied without modifications, may yield different results for the same MOTS. Moreover, no mathematical or physical argument is provided for the validity of the conformal approach.

In this work we also aim to give a more rigorous treatment of the conformal decomposition method in the context of
black hole spin definition.
In particular, we explain how one can make the definition completly insensitive to various
gauge choices which must be made when describing the geometry of MOTS's.
 
The application of the framework is confined to two--dimensional surfaces of
$S^2$ topology and cannot be generalized to other dimensions and topologies.

\section{Conformal decomposition and transformations}
On any oriented manifold of topology $S^2$, equipped with a positive definite metric $q$, 
there exist coordinates $(\theta,\varphi)$ for which the metric takes globally
the conformally spherical form
\begin{displaymath}
q = F(\dd\theta^2+\sin^2\theta\,\dd\varphi^2), 
\end{displaymath}
$F(\theta,\varphi)$ being a positive function, and preserving the orientation \cite{czuchry-2004-70,dfn-book}. (For a modern approach to constructing such systems
using the Ricci flow techniques see \cite{chow-1991-33,hamilton-1988-71}, the latter also in \cite{cao-book}.)
The choice of the conformally spherical coordinate system (CSCS) is by no means  unique. However any two such systems are related to each other by a global conformal transformation of the
``round'' sphere metric $q_0 = \dd\theta^2 + \sin^2\theta\,\dd\varphi^2$. 
Such transformations are known to constitute a six parameter group, isomorphic to the connected
component of $SO(1,3)$ 
\cite{sharpe-booke}. In the context of the Riemann sphere it is also called the M\"obius group.

The group consists of the $SO(3)$ subgroup of ``standard'' rotations,
preserving $q_0$, and so--called proper conformal transformations \cite{sharpe-booke}.
Its action on $\Delta$ is generated by six vector fields, three of them generating  the rotations about the three orthogonal axes
\begin{eqnarray}
\phi_1 &=& -\sin\varphi\,\partial_\theta - \cotan\theta\,\cos\varphi\,\partial_\varphi\nonumber\\
\phi_2 &=& \cos\varphi\,\partial_\theta - \cotan\theta\,\sin\varphi\,\partial_\varphi\nonumber\\
\phi_3 &=& \partial_\varphi\nonumber
\end{eqnarray}
and three generating the proper conformal transformations along the three axes
\begin{eqnarray}
\xi_1 &=& -\cos\theta\,\cos\varphi\,\partial_\theta + \frac{\sin\varphi}{\sin\theta}\,\partial_\varphi \nonumber\\
\xi_2 &=& -\cos\theta\sin\varphi\,\partial_\theta - \frac{\cos\varphi}{\sin\theta}\,\partial_\varphi \nonumber\\
\xi_3 &=& \sin\theta\,\partial_\theta\nonumber. 
\end{eqnarray}
Any combination of the form of $n_i\,\phi_i$ with $n_i \,n_j \,\delta^{ij}=1$ is an axial vector field in the terminology of the previous section, while
 no linear combination of $\xi_i$'s is axial. 
Together these vector fields constitute the Lie algebra of $so(1,3)$ with commutation relations
 \begin{eqnarray}
 \left[\phi_i,\phi_j\right] &=& -\epsilon_{ijk}\,\phi_k \label{eqcomphiphi}\\
 \left[\xi_i,\xi_j\right] &=& \epsilon_{ijk}\,\phi_k \label{eqcomxixi}\\
 \left[\xi_i,\phi_j\right] &=& \left[\phi_i,\xi_j\right] = -\epsilon_{ijk}\,\xi_k
 \label{eqcomxiphi}
 \end{eqnarray}
($\epsilon_{ijk}$ is the Levi--Civitt\`a antisymmetric symbol). 
The vector fields have been defined in the language of the CSCS and therefore any
transformation of the coordinates affects them as well. Namely, the action of infinitesimal transformations  $\dot\theta = \Lie_P\, \theta$, $\dot\varphi = \Lie_P\, \varphi$ is
given by
\begin{eqnarray}
\dot \phi_i &=& \left[P,\phi_i\right] \label{eqphitrans}\\
\dot \xi_i &=& \left[P,\xi_i\right] \label{eqxitrans}
\end{eqnarray}
(the vector field $P$ is any linear combination of $\phi_i$ and $\xi_i$).
It follows easily that under the rotation given by a matrix $\Lambda\UD{j}{i} \in SO(3)$ 
the vector fields defined above transform according to
\begin{eqnarray}
\widetilde\phi_i &=& \Lambda\UD{j}{i}\,\phi_j \label{eqphirotations}\\
\widetilde \xi_i &=& \Lambda\UD{j}{i}\,\xi_j \label{eqxirotations}.
\end{eqnarray}
The integrated action of a proper conformal transformation generated by $n_i\,\xi_i$, 
with $n_i\,n_j\,\delta^{ij} = 1$, is given by
\begin{widetext} 
\begin{eqnarray}
\widetilde\phi_i &=& n_i\,(n_k\,\phi_k) + \cosh\lambda \left(\phi_i - n_i\,(n_k\,\phi_k)\right)
+ \sinh\lambda \,(\epsilon_{ijk}\,n_j\,\xi_k) \label{eqphipropers}\\
\widetilde\xi_i &=& n_i\, (n_k\,\xi_k) + \cosh\lambda \left(\xi_i - n_i\,(n_k\,\xi_k)\right)
- \sinh\lambda \,(\epsilon_{ijk}\,n_j\,\phi_k) \label{eqxipropers},
\end{eqnarray}
\end{widetext}
where $\lambda$ is the group additive parameter.
Note that in our convention the action of the conformal group is passive, i.e. it 
acts \emph{only} on the coordinate systems  and \emph{not} on $\omega$ or $q$.

 We introduce two triples of integrals
 \begin{eqnarray}
 J_i &=& -\frac{1}{8\pi G} \int_\Delta \omega(\phi_i)\,\epsilon \label{eqintj}\\
 K_i &=& -\frac{1}{8\pi G} \int_\Delta \omega(\xi_i)\,\epsilon \label{eqintk}.
 \end{eqnarray}  
From (\ref{eqxirotations}--\ref{eqphirotations}) and (\ref{eqintj}--\ref{eqintk}) we see that under the action of the rotation subgroup $SO(3)$ these triples transform like standard three--dimensional vectors, so
it is legitimate to consider them as vectors in a three--dimensional 
Euclidean space and denote by $\vec J$ and $\vec K$. 

It would be tempting to use the length of $\vec J$ as a measure
of angular momentum (as was effectively done in \cite{caudill-2006-74}), but we must keep in mind that from the beginning we have a freedom
of choosing the CSCS, which affects the values of the integrals (\ref{eqintj}) and (\ref{eqintk}). 
Although  $|\vec J|$ is invariant under rotations, the proper conformal
transformations in general mix $\vec J$ and $\vec K$ and do not preserve their norms. Namely, 
the integrated version of the proper conformal transformations (\ref{eqxipropers}--\ref{eqphipropers}), combined with (\ref{eqintj}) and (\ref{eqintk}) yields
\begin{eqnarray}
\vec J' &=& \gamma\left(\vec J + \vec\beta \times \vec K\right) - \frac{\gamma^2}{\gamma+1}
\vec\beta (\vec\beta \cdot \vec J) \label{eqjtrans}\\
\vec K' &=& \gamma\left(\vec K - \vec\beta \times \vec J\right) - \frac{\gamma^2}{\gamma+1} 
\vec \beta (\vec\beta \cdot \vec K) \label{eqktrans},
\end{eqnarray}
where we have introduced for convenience 
\begin{eqnarray}
\vec\beta &=& \tanh\lambda\cdot\vec n,\qquad |\vec \beta|<1 \nonumber\\
 \gamma &=& (1-\vec\beta^2)^{-1/2} \nonumber   
\end{eqnarray}
(scalar and vector products are defined in a standard way).

Note that equations (\ref{eqjtrans}) and (\ref{eqktrans}) are exactly the same as the transformations laws 
for the electric and magnetic field vectors $\vec E$ and $\vec B$ under the
Lorentz boosts \cite{landaulifshitz}. Since the action of
rotations is also identical, we conclude that under all (orthochronal) $SO(1,3)$ transformations  $\vec J$ and $\vec K$
transform exactly like the electric and magnetic field.  We will explore this 
 unexpected analogy in the rest of the paper, as for now noting only that
 that these transformations have two well--known polynomial invariants of  second degree \cite{landaulifshitz} 
\begin{eqnarray}
A &=& |\vec J|^2 - |\vec K| ^2 \label{eqa}\\
B &=& \vec K \cdot \vec J. \label{eqb}
\end{eqnarray}
As invariants, they do not depend on the initial choice of the CSCS and
 are therefore well--defined quantities on any MOTS. 

\section{Definition of the axial vector and angular momentum}
If $\vec J$ and $\vec K$ are not parallel in a given CSCS and the invariants $A$ and $B$ do not
vanish simultaneously, there exists a proper conformal transformation which 
makes them parallel or causes one of them to vanish \cite{landaulifshitz}. 
It is given by (\ref{eqjtrans}) and (\ref{eqktrans}) with
\begin{equation}
\vec\beta = \beta\frac{\vec J \times \vec K}{|\vec J\times\vec K|}, \nonumber
\end{equation}
where $\beta$ is the only root of
\begin{equation}
1 - \frac{|\vec J|^2 + |\vec K|^2}{|\vec J\times\vec K|}\beta + \beta^2 = 0 \nonumber
\end{equation}
satisfying  $0 < \beta < 1$. Once it has been applied, the only transformations preserving the parallelness of $\vec J$ and $\vec K$ are
the rotations and the proper conformal transformations with $\vec\beta$ parallel to 
$\vec J$ and $\vec K$. It is straightforward to check that they do not affect $|\vec J|$ and $|\vec K|$. Moreover, 
the axial vector field $\phi$ given by 
\begin{equation}
\phi = \frac{J_i}{|\vec J|}\,\phi_i = \pm \frac{K_i}{|\vec K|}\,\phi_i
\end{equation}
is also invariant and substituted into (\ref{eqj}) yields exactly $|\vec J|$. All these observations justify the following definition:
\begin{definition}
For any MOTS for which $A^2 + B^2 > 0$  we define
the value of angular momentum as
\begin{equation}
J = |\vec J| \nonumber
\end{equation}
calculated in any conformally spherical coordinate system in which $\vec J \times
\vec K = 0$. The corresponding axial vector
field $\phi$ is defined in the same coordinates as  
\begin{equation}
\phi = \frac{J_i}{|\vec J|}\,\phi_i \nonumber
\end{equation}
if  $\vec J \neq 0$, or
\begin{equation}
\phi = \frac{K_i}{|\vec K|}\,\phi_i \nonumber
\end{equation} otherwise. 
\end{definition}
The definition obviously satisfies the requirements 2 and 4. We will now prove that $J$ and
$\phi$  
coincide with the ``standard'' ones on an axisymmetric MOTS (condition 3).

One can easily prove that if the MOTS admits an axial symmetry of both the metric and the rotation one--form,
i.e. there exists an axial vector $X$ such that $\Lie_X \omega = 0$,
$\Lie_X q=0$, 
then it is possible to find an adapted CSCS $(\theta,\varphi)$ in which
 $X = \partial_\varphi$ and consequently the metric and the rotation form
take a particularly simple form
\begin{eqnarray}
q&=&F(\theta)(\dd\theta^2 + \sin^2 \theta\,\dd\varphi^2)\nonumber\\
\omega &=& \omega_\theta(\theta)\,\dd\theta+\omega_\varphi(\theta)\,\dd\varphi\label{eqomegaaxisym}.
\end{eqnarray}
In this case all integrals in (\ref{eqintj}) and (\ref{eqintk}) vanish except
\begin{eqnarray}
J_3 &=& -\frac{1}{8\pi G}\int_\Delta \omega_\varphi(\theta)\,\epsilon \label{eqj3}\\
K_3 &=& -\frac{1}{8\pi G}\int_\Delta \sin\theta\, \omega_\theta(\theta)\,\epsilon \nonumber
\end{eqnarray}
and we see that $\vec J$ and $\vec K$ are both parallel to the symmetry axis and in consequence $J =
|\vec J|$. The symmetry vector $X$ is now equal to the axial vector field $\partial_\varphi$, which in turn is equal up to 
 sign to $\frac{J_i}{|\vec J|}\,\phi_i$ and $\frac{K_i}{|\vec K|}\,\phi_i$,
whenever the latter two are defined. 
Consequently $J = \pm J_3$, which is again equal up to sign to $J_X$, as expected. This completes the proof.

The value of $J$ can be conveniently expressed in terms of the two invariants $A$ and $B$.
Namely, assuming that $\vec J$ and $\vec K$ are parallel we can solve (\ref{eqa}) and (\ref{eqb}) for
$J$ obtaining
\begin{equation}
J = \sqrt{\frac{A + \sqrt{A^2+4B^2}}{2}}.  
 \label{eqjdef2}
\end{equation}
We may regard (\ref{eqjdef2}) as another definition of the angular momentum which,
contrary to the previous one, is manifestly $SO(1,3)$--invariant and therefore applicable to \emph{any} 
conformally spherical coordinates. This fact makes it much more useful from
the computational point of view.

Note that formula (\ref{eqjdef2}) is perfectly valid even in the ``plane wave'' case, when both $A$ and $B$ vanish. Thus the value of angular momentum 
continues to be well--defined despite the fact that the axial vector field is not. 
The inapplicability of the previous, geometric definition to the ``plane wave'' MOTS's seems less surprising 
if we consider that in that case (\ref{eqjdef2}) yields identically 0.   
Obviously if angular momentum vanishes, the axis of rotation, and consequently the axial vector field
$\phi$, is undefined.

 This introduces a slight complication, because
 without a vector field one cannot put (\ref{eqjdef2}) 
in the form of (\ref{eqj}), which is crucial to apply the Hamiltonian formulation
in order to prove that $J$ is the generator of the horizon rotations. Therefore if $A=B=0$, formula (\ref{eqjdef2}) for angular momentum can only be justified by a
 continuity argument.

 \section{Gauge invariance}
 As we mentioned in the introduction, renormalizing the null normals or changing the foliation of an
 isolated horizon results in 
 adding the gradient of a complete function on $\Delta$ to the rotation form
  \cite{korzynski-2006-74,lewandowski-2006-23}.
 If the vector field $\phi$ preserves the area form, formula (\ref{eqjdef2}) is insensitive to
 such changes \cite{gourgoulhon-2005-72}. In our construction however the axial vector field 
 needn't be divergence--free  and adding a gradient to the rotation form 
 usually affects the value of
  $J$. 
 
 One could possibly modify the vector field $\phi$ in such a way that it becomes 
 divergence--free, for example by an appropriate pointwise rescaling. Namely, one can verify that
  vector field given $\widehat\phi$ by
 \begin{eqnarray}
 \widehat\phi &=& \frac{C(\theta)}{F(\theta,\varphi)}\,\phi \nonumber\\
 C(\theta)&=&\frac{1}{2\pi}\int_{0}^{2\pi} \,F(\theta,\varphi)\,\dd \varphi, \nonumber
 \end{eqnarray}
 calculated in the CSCS with $\vec J\times\vec K = 0$,  is both axial and divergence--free
 (the coordinate system is assumed to have been rotated so that
 $\phi = \partial_\varphi$).
This rescaling however makes the definition of angular momentum more complicated. In particular, since $\widehat\phi$ 
needn't be a combination of the M\"obius group generators, we loose the simple
  way to evaluate $J$ offered by equation (\ref{eqjdef2}). 
  Therefore we will follow another path here.
 
 Another way around the problem of gauge dependence would be to 
 take into account only the gauge invariant part of $\omega$ in the definition. 
 This can be achieved easily and without violating (\ref{eqj}) by appropriate fixing of the 
 normalization 
 of the null normals.
Namely, one should require that in the Hodge decomposition $\omega = \star\,\dd f + \dd g$ the
gradient part $\dd g$ vanishes, or equivalently that
\begin{equation}
\dd\star\omega = 0. \label{eqgaugefix} 
\end{equation}
 It is straightforward to verify that it is always possible to gauge $l^\mu$ and 
$k^\mu$ in such a way that (\ref{eqgaugefix}) holds.
Since the Hodge operator acting on one--forms is conformally invariant,
it does not matter whether it is taken with respect to the original metric $q$ or the ``round'' one.
Note that on an axisymmetric horizon (\ref{eqgaugefix}) is equivalent to setting the first
term in (\ref{eqomegaaxisym}) to 0. This term does not contribute to the value of $J$, 
as we can see in (\ref{eqj3}), so  
  the proposed gauge fixing doesn't matter in axisymmetric surfaces. In particular it
  does not spoil the compliance of the presented definition with condition 3.

Fixing the null normals by (\ref{eqgaugefix}) has yet another nice consequence.
In a Kerr or Kerr--Newman horizon, or in any axisymmetric horizon, if we change the foliation from the standard one to one with
 arbitrarily tilted or waived leaves, the rotation form will also acquire a gradient
 of a function of both coordinates $h(\theta,\varphi)$. This destroys the rotation invariance 
 of $\omega$ with respect
 to $\partial_\varphi$ and potentially changes the value of $J$. However condition (\ref{eqgaugefix})
  imposes renormalization
 of the null normals which exactly cancels with $\dd h$. Thus  
 equation (\ref{eqjdef2}) with the described gauge fixing 
  yields the expected value of angular momentum when applied
 to any section of the Kerr or Kerr--Newman horizon.   
   
 Of course in practical calculations it is not necessary to actually rescale the null normals.
 It suffices to perform the Hodge decomposition of the rotation form and substitute only
 $\star \,\dd f$ as the rotation form into any of the definitions.

 \section{Properties}
 We will now briefly discuss several properties of the proposed definition of angular momentum.
 First, if $\vec K$ and $\vec J$ turn out to be perpendicular in a particular CSCS, (\ref{eqjdef2})
  can be 
 simplified to
 \begin{equation}
J = \left\{\begin{array}{ll} \sqrt{|\vec J|^2 - |\vec K|^2} \quad \textrm{    if  $|\vec J| > |\vec K|$,} \\ \\ 0 \quad\textrm{   otherwise.}\end{array}\right. 
\end{equation}
Thus the relationship between $J$ and $\vec J$ turns out to be non--differentiable.
Moreover, angular momentum vanishes identically whenever $|\vec J| \le |\vec K|$. 
This is slightly surprising, but perfectly consistent with the
first definition of $J$. Namely, if $|\vec K|>|\vec J|$ or $|\vec K|<|\vec J|$, one can pass to the CSCS with $\vec J\times\vec K = 0$ where it turns out that $\vec J=0$ or $\vec K = 0$ respectively.
The intermediate case of both vectors being of equal norms is exactly the ``plane wave'' case when we cannot 
apply the first definition.

Note that something similar happens even if $\vec J$ and $\vec K$ are not perpendicular,
though not parallel either:
if we increase the length of $\vec K$ keeping $\vec J$ fixed, the value of angular momentum
 tends to 0, although no discontinuity of the first derivative is present.   

None of these peculiarities appears in axisymmetric horizons. 
As we noted in the previous section, imposing (\ref{eqgaugefix}) ensures that both vectors are always
parallel and therefore changing $|\vec K|$ does not affect $J$.

\section{Physical motivation behind the parallelness condition}

One may ask whether there exists a physical motivation behind the requirement of $\vec J$ and $\vec K$ being parallel, other than the correct behavior of $J$ in known and obvious cases.
In fact, a simple analogy may be drawn between the MOTS's and  relativistic systems of non--interacting particles in a Minkowski
background.
If $x_N^\mu$ denotes the spacetime coordinates of particle $N$ and 
$p_N^\mu$ its four--momentum, we can define two vectors
\begin{eqnarray}
\mathcal{K}_i &=& \sum_N \, (p^i_N \,x^0_N - p^0_N\,x^i_N) \nonumber\\ 
\mathcal{J}_i &=& \sum_N \,\epsilon_{ijk}\,x^j_N\,p^k_N, \nonumber
\end{eqnarray}
where the summation over all particles is done at a given instant of coordinate time $x^0$.
These vectors are the non--vanishing components of the antisymmetric angular momentum
four--tensor \cite{schwartz-book}. 
They generate the action of the Lorentz group on the phase space in
the same way (\ref{eqintj}--\ref{eqintk}) generate the action of the M\"obius group on a MOTS.
Assume that in a given reference frame we have shifted the origin to the
 momentary position of system's centroid calculated in that frame, i.e. we have
\begin{equation}
\sum_N p^0_N\,x^i_N = 0. \nonumber
\end{equation}
Now one can verify that 
 if  $\mathcal{\vec J}$ is parallel to $\mathcal{\vec K}$, then the norm of $\mathcal{\vec J}$ is
  equal to the value of system's total angular momentum evaluated in the center--of--mass frame.
  The latter is the ``true'' intrinsic angular momentum given by the norm of
  the Pauli--Luba\'nski vector \cite{rindler-book}. 
 
   \section{Summary and further developments}
We will now briefly summarize the main result of the paper.
Basing on the action of the conformal group on a MOTS, 
we have presented a new definition of angular momentum $J$, applicable to
both axisymmetric and non--symmetric surfaces, along with a 
construction of the corresponding axial vector field whose flow
is generated by the angular momentum as a Hamiltonian generator. The construction
of the vector field works for all surfaces except a narrow class in which the value
of angular momentum vanishes. 

Having noticed that the value of $J$ depends on the choice of null normals,
we have proposed a simple gauge condition which removes this ambiguity. 
The resulting definition yields the expected answer when applied to axisymmetric horizons,
even when the choice of foliation or normalization is not compatible with the symmetry.

In the next paper \cite{korzynski-inprep}, apart from discussing another proposition for the angular momentum definition, we will derive the angular momentum flux law and the 
first law of black hole mechanics in non--axisymmetric horizons. 

\begin{acknowledgements}
The author would like to thank Jerzy Kijowski, Tomasz Paw\l{}owski,  Jerzy Lewandowski, Jacek Jezierski, Micha\l{} Godli\'nski and Badri Krishnan for valuable discussions and comments. 
The author is especially grateful to 
 Lars Andersson for discussions and invitation to the MPI. The work was supported by the Polish Ministry of Science and Higher Education as a research project (grant No. N202 016 31/0645) from the means for years 2006 and 2007, and by the Max Planck Institute in Golm. 
\end{acknowledgements}

\bibliography{moments}

\end{document}